\documentclass[conference]{IEEEtran}
\IEEEoverridecommandlockouts

\usepackage{tikz}
\usepackage{textcomp}
\usepackage{hyperref}
\usepackage{lipsum}

\newcommand\copyrighttext{%
  \footnotesize \textcopyright 2023 IEEE. Personal use of this material is permitted. Permission from IEEE must be obtained for all other uses, in any current or future media, including reprinting/republishing this material for advertising or promotional purposes, creating new collective works, for resale or redistribution to servers or lists, or reuse of any copyrighted component of this work in other works.
  DOI: \href{https://doi.org/10.1109/OSMSES58477.2023.10089686       }{10.1109/OSMSES58477.2023.10089686}}
\newcommand\copyrightnotice{%
\begin{tikzpicture}[remember picture,overlay]
\node[anchor=south,yshift=10pt] at (current page.south) {\fbox{\parbox{\dimexpr\textwidth-\fboxsep-\fboxrule\relax}{\copyrighttext}}};
\end{tikzpicture}%
}
\usepackage{cite}
\usepackage{amsmath,amssymb,amsfonts}
\usepackage{algorithmic}
\usepackage{graphicx}
\usepackage{textcomp}
\usepackage{xcolor}
\usepackage{caption}
\usepackage{subcaption}
\usepackage{enumitem}
\usepackage{url}

\def\BibTeX{{\rm B\kern-.05em{\sc i\kern-.025em b}\kern-.08em
    T\kern-.1667em\lower.7ex\hbox{E}\kern-.125emX}}
\begin{document}

\title{Load and generation time series for German federal states: Static vs. dynamic regionalization factors}

%

\author{\IEEEauthorblockN{1\textsuperscript{st} Madeleine Sundblad}
\IEEEauthorblockA{\textit{Department of Sustainable Systems Engineering (INATECH),} \\
\textit{University of Freiburg} 79110, Freiburg, Germany} 
\and
\IEEEauthorblockN{2\textsuperscript{nd} Tim F\"urmann}
\IEEEauthorblockA{\textit{Department of Sustainable Systems Engineering (INATECH),} \\
\textit{University of Freiburg} 79110, Freiburg, Germany}
\and
\IEEEauthorblockN{3\textsuperscript{rd} Anke Weidlich}
\IEEEauthorblockA{\textit{Department of Sustainable Systems Engineering (INATECH),} \\
\textit{University of Freiburg} 79110, Freiburg, Germany}
\and
\IEEEauthorblockN{4\textsuperscript{th} Mirko Sch\"afer}
\IEEEauthorblockA{\textit{Department of Sustainable Systems Engineering (INATECH),} \\
\textit{University of Freiburg} 79110, Freiburg, Germany\\
mirko.schaefer@inatech.uni-freiburg.de}
}
\maketitle
\copyrightnotice

\begin{abstract}
Electricity generation and demand time series often are only available on a national scale. In this contribution, we derive regionalization factors to allocate publicly available national generation and demand time series for Germany to the federal-state level. We compare two different types of regionalization approaches: Static factors are based on the regional distribution of capacities or population and GPD, whereas dynamic factors take plant-specific generation time series, regionally resolved weather patterns or compositions of different load profiles into account. We observe that dynamic regionalization factors show significant temporal variability, emphasizing the limitations of static regionalization factors for a spatio-temporally more detailed representation of power system time series.
\end{abstract}

\section{Introduction}
System data with a high temporal and spatial resolution is of interest for energy system analysis and modelling, but also for monitoring regional indicators like the degree of self-sufficiency or local grid emission factors. One main area of research is the collection and modelling of data which serves as input for detailed energy system models. Examples include renewable generation time series derived from weather data and wind turbine or photovoltaic panel models~\cite{jung2022,atlite,victoria2019}, data sets describing renewable energy potentials through land availability~\cite{risch2022}, or historical and projected regionalized load time series~\cite{kuehnbach2021,demandregio,buettner2022}. Nevertheless, for model validation as well as system analysis and monitoring, spatio-temporally resolved data representing the actual system is needed. For Germany, historical and close to real-time data in general is provided by system operators for each control area and on a national level. The four German transmission system operators provide, for instance, electricity demand or per type generation time series with a temporal resolution of 15 minutes. A central provider of power system data is the electricity market platform "SMARD" hosted by the Federal Network Agency ("Bundesnetzagentur")\cite{SMARD}. On an European scale, the European Network of Transmission System Operators for electricity (ENTSO-E) publishes a wide range of electricity system data and time series through the ENTSO-E Transparency Page~\cite{transparency}. Using this data as input, different institutions visualize (if necessary preprocessed) energy system data, thus providing information and facilitating access to electricity system data for the public~\cite{agorameter,energycharts}. Electricity system data from ENTSO-E also serves as an input to derive further system information like, for instance, hourly generation- and consumption based emission intensity signals for European countries~\cite{electricitymap,tranberg}.
All these applications share the common restriction to be limited by the spatial and temporal scale of the original data provided by the system operators. Data with a higher spatial resolution has to be either collected from additional sources like, for instance, load data from the numerous distribution grid operators, or has to apply regionalization processes. Examples include the load regionalization methods derived from the DemandRegio project~\cite{demandregio}, or the simplified capacity-based regionalization process applied to derive electricity generation time series for the federal state Nordrhein-Westfalen~\cite{energieatlas}.

We study this problem of representing electricity system time series on a regional level through the calculation of regionalization factors for federal states in Germany. As input, we assume load and per type generation time series on a national level as provided by ENTSO-E or SMARD. Regionalization factors distribute this nationally aggregated information to the different federal states. We show that regionalization factors, which are based on static information  (monthly or annual time scale) like capacity distributions or population statistics, neglect the significant temporal variability represented by dynamic regionalization factors, which take into account time-dependent information (hourly time scale). These results indicate that simplified static regionalization methods provide information with very limited accuracy, and motivate efforts to apply more elaborated dynamic regionalization methods.

The article is structured as follows: Section~\ref{sec:data} discusses both the underlying data sources and the methods to derive static and dynamic regionalization factors for federal states in Germany. The subsequent section~\ref{sec:results} presents results with a focus on the comparison between the static and dynamic approaches. Section~\ref{sec:conclusion} concludes the article.

\section{Data regionalization}\label{sec:data}
In the following, we will assume the overall time frame of one year (2021), with all time series available with at least hourly resolution. As input data, we use electricity demand time series $d(t)$ and per type generation series $g^{s}(t)$ on the national level. Here the technology index $s$ denotes the generation type, i.e. "Fossil Hard Coal", "Nuclear", "Wind Onshore" etc. These time series have to be regionalized to determine the corresponding quantities on the federal-state level:
\begin{equation}
\begin{aligned}
d_{n}(t) = r^{d}_{n}(t)\cdot d(t)\quad,\quad
g_{n}^{s}(t)=r^{s}_{n}(t)\cdot g^{s}(t)~.
\end{aligned}
\end{equation}
Here the index $n$ denotes the region, i.e. the individual federal state. We consider both static regionalization factors $r_{n}(t)=r_{n}$ and dynamic regionalization factors $r_{n}(t)$. Static factors can be derived from time-aggregated quantities like the share of generation capacity per region or population. Dynamic factors need to take some time-dependent signal into account, for instance different load profiles for different regions, or spatio-temporal weather patterns for the renewable generation time series.

For this study, data for the national electricity demand and per type generation time series has been taken from the ENTSO-E Transparency Platform (actual load and actual generation per production type~\cite{ENTSOE_generation,ENTSOE_load}). All data has been resampled to hourly resolution. In the following, regionalization factors for the per type generation time series are derived for the technology categories as classified by ENTSO-E (see Table~\ref{tab:technologies}). 
The methodology can be adapted to similar time series through a corresponding classification of the technology indices.
\subsection{Static regionalization factors}\label{sec:static}
Static regionalization factors for all generation time series are derived from the ratio of installed capacity per production type in a state to the total installed capacity per production type in Germany:
\begin{equation}
\begin{aligned}
r^{s}_{n}=\frac{G^{s}_{n}}{\sum_n G^{s}_{n}}~.
\label{eq:reg_stat_gen}
\end{aligned}
\end{equation}
Here $G_{n}^{s}$ denotes the installed capacity of type $s$ in federal state $n$. The underlying data is collected from the German Federal Network Agency (Bundesnetzagentur) power plant list~\cite{BNetzA}. This regularly updated list identifies and describes the net and gross capacity, production type, location and working status of the individual power plants in Germany. Power generation units are identified using two codes, the MaStR number and the power plant number of the Federal Network Agency. For 2021, this dataset represents in total 212.46~GW of installed capacity, neglecting small units with capacities below 10~MW. The renewable generation is not separated by unit but is provided as the total sum of capacity per technology type per state. Offshore wind turbines are not assigned to any federal state in the power plant list and have been manually assigned based on information from~\cite{WindGuard}. Based on Eq.~\ref{eq:reg_stat_gen}, a total of 224 distribution factors are created (16 states and 14 ENTSO-E production type categories as listed in Table~\ref{tab:technologies}).
\begin{table}[!t]
\centering
\caption{Mapping of production type classifications in the Federal Network Agency (BNetzA) power plant list to ENTSO-E technology classifiers}
\label{tab:technologies}
\begin{tabular}{lr} 
\hline
ENTSO-E technology type & BNetzA technologies\\
\hline
Fossil Oil & Mineral\"olprodukte\\
Fossil Gas & Erdgas, Grubengas\\
Fossil Hard Coal & Steinkohle\\
Fossil Brown coal/ & Braunkohle\\
Lignite & \\
Biomass & Biomasse, Deponiegas, Kl\"argas\\
Waste & Abfall\\
Other & Sonstige Energietr\"ager (nicht erneuerbar),\\ & Batteriespeicher, W\"arme, Mehrere\\
& Energietr\"ager, Sonst. Speichertechnologien,\\
& Andere Gase, Kl\"arschlamm, Unbekannter\\
& Energietr\"ager\\
Hydro Run-of-river & Wasser, Laufwasser,\\
and poundage& Speicherwasse (ohne Pumpspeicher)\\ 
Hydro Pumped Storage & Pumpspeicher\\
Nuclear & Kernenergie\\
Other renewable & Sonstige Energietr\"ager (erneuerbar)\\
Solar & Solare Strahlungsenergie\\
Wind Offshore & Windenergie (Offshore-Anlage)\\
Wind Onshore & Windenergie (Onshore-Anlage)\\
\hline
\end{tabular}
\end{table}
The static regionalization factors for the ENTSO-E actual load time series are based on population and gross domestic product data. For each federal state, the national hourly demand time-series is distributed proportionally to a weighted share of 60\,\% GDP and 40\,\% population. This weighting has been suggested in~\cite{PyPSA-Eur} based on a linear regression and has also been used in~\cite{Unnewehr2022}.

\subsection{Dynamic regionalization factors}\label{sec:dynamic}
\begin{figure*}
\centering
\includegraphics[width=0.8\textwidth]{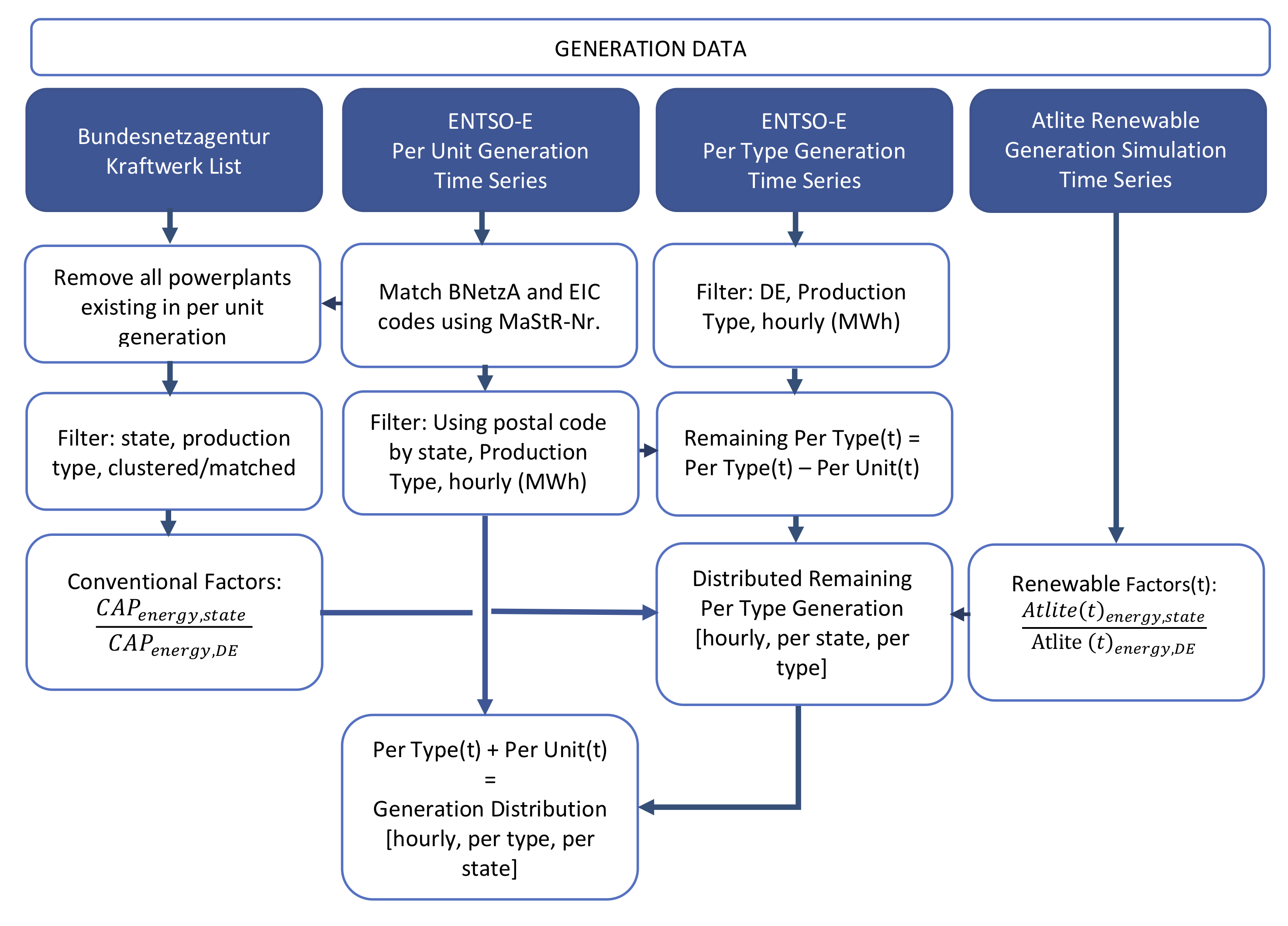}
\caption{Process diagram for the calculation of dynamic regionalization factors for generation time series.}
\label{fig:dynamic}
\end{figure*}

The dynamic regionalization process for generation data is visualized in Fig.~\ref{fig:dynamic}.
ENTSO-E publishes per unit generation time series for generation units with more than 100 MW generation capacity. For Germany, this accounts for 182 generation units, providing 40\% of total generation. Each generation unit covered by this data set has been categorized according to the associated ENTSO-E technology category and matched with the corresponding entry in the Federal Network Agency entry. A part of the per type generation thus can be distributed to the Federal States according to the location of the generation units covered by the per generation time series. The remaining per type generation data not represented by the per unit generation time series has been regionalized using static factors as described in Sec.~\ref{sec:static}, only that the Federal Network Agency power plant list has been reduced to the units for which generation time series are not published by ENTSO-E. 

\begin{figure}
\centering
\includegraphics[width=0.5\textwidth]{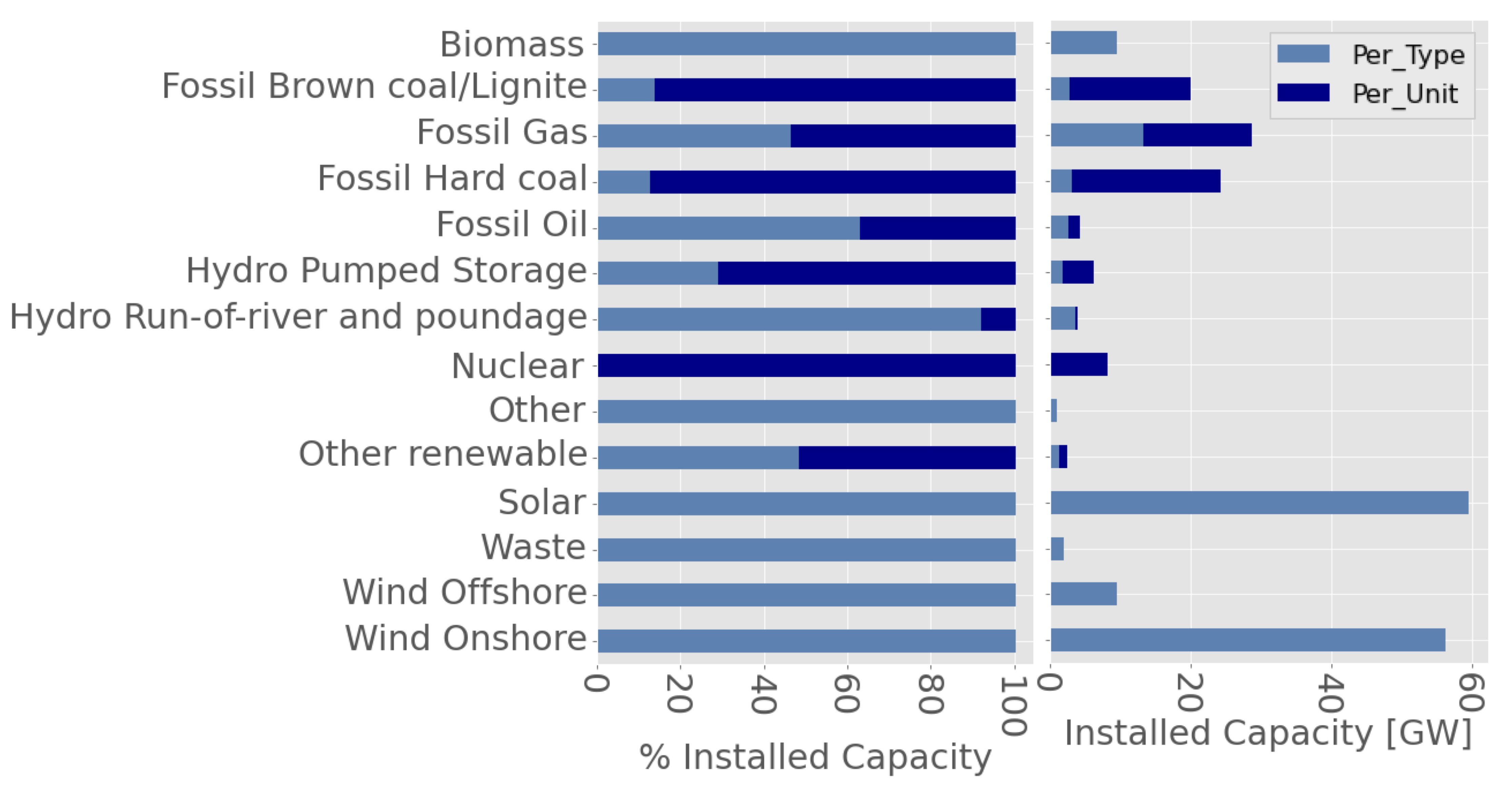}
\caption{Amount of generation capacity per technology covered by the per unit generation time series data from ENTSO-E. Left-hand side shows the relative share of the covered generation, right-hand side shows corresponding capacities in GW.}
\label{fig:per_generation}
\end{figure}

Figure~\ref{fig:per_generation} shows for each of the ENTSO-E technology types the amount of capacity covered by the per unit generation time series, both as shares and absolute values. A large part of the generation units from Fossil Hard Coal, Brown Coal/Lignite, and Nuclear have generation capacities larger than 100 MW and thus are covered by the per unit generation data set. For these technologies, the per unit time series contribute 80-100 \% of the total installed capacity of their respective energy type. Due to a large number of smaller generation units, for Fossil Gas only approximately 50 \% of the generation is covered by per unit time series. 
Dynamic regionalization factors for renewable generation from Wind (Onshore and Offshore) and Solar for each federal state are calculated based on weather data and installed renewable capacities through the simulation tool \emph{atlite}~\cite{atlite}. The weather data for the year 2021 was obtained from the ECMWF Reanalysis v5 (ERA5) dataset ~\cite{hersbach2020era5} and the information about location, capacity, and type of the individual renewable power plants was taken from the German Marktstammdatenregister (MaStR, central register for installation data)~\cite{MaStR}. In order to match the renewable power plant locations to the federal states, shapefiles were obtained from the natural earth free vector and raster map dataset\cite{NaturalEarth}. 
The general workflow of the calculation of the dynamic regionalization factors for wind and solar power generation consists of the following steps.
\begin{enumerate}
    \item The renewable power plant data from the central register for installation data was filtered regarding location, capacity and technology. Dataset entries without geographic information were neglected. The remaining renewable power plants were mapped to the federal states by comparing if their location was within the borders of the corresponding federal state. In the case of Offshore Wind, the nearest grid entry point was chosen via the haversine formula and the ENTSO-E grid map.
    \item \emph{atlite} was used to calculate the renewable generation for each renewable power plant based on its location, its capacity and the weather data taken from the ERA5 dataset. A cutout of Germany based on its geometric boundaries was created matching the grid cells of the ERA5 data and the underlying coordinate system of the German map. The renewable generation was then calculated using \emph{atlite}'s conversion function~\cite{atlite} for solar photovoltaic, onshore wind and offshore wind. The used generation models for renewable generation were the following:\vspace{6pt}\\
    \begin{tabular}{lp{5.5cm}}
        Solar: & CSI panel with a slope of 30° and an azimuth angle of 180°\\
        Onshore: & Vestas V112 turbine with 3~MW rated capacity\\
        Offshore: & NREL reference turbine with 5~MW rated capacity\\
    \end{tabular}\vspace{6pt}\\
    The resulting renewable generation per type and location was then aggregated to the federal-state level.
    \item The dynamic regionalization factors per type and federal state $r^{s}_{n}(t)$ are  calculated by taking the fraction between renewable generation per type and location aggregated to the federal state level $g^{s}_{n}(t)$ and the overall renewable generation per type for Germany $\sum_{n} g^{s}_{n}(t)$:\vspace{3pt}
    \begin{center}
        \begin{math}
            r^{s}_{n}(t)=\frac{g^{s}_{n}(t)}{\sum_{n} g^{s}_{n}(t)}~.
        \end{math}
    \end{center}
\end{enumerate}
\vspace{12pt}

Dynamic regionalization factors for electricity demand are derived from an hourly resolved demand data set for all NUTS3-regions in Germany for 2015~\cite{kuehnbach2021}. Hourly values have been aggregated for each Federal State, and the entire data set temporally shifted to match the days of the week of the year 2021. 

\section{Results}\label{sec:results}
\begin{figure*}[htbp]
\centering
\includegraphics[width=\textwidth]{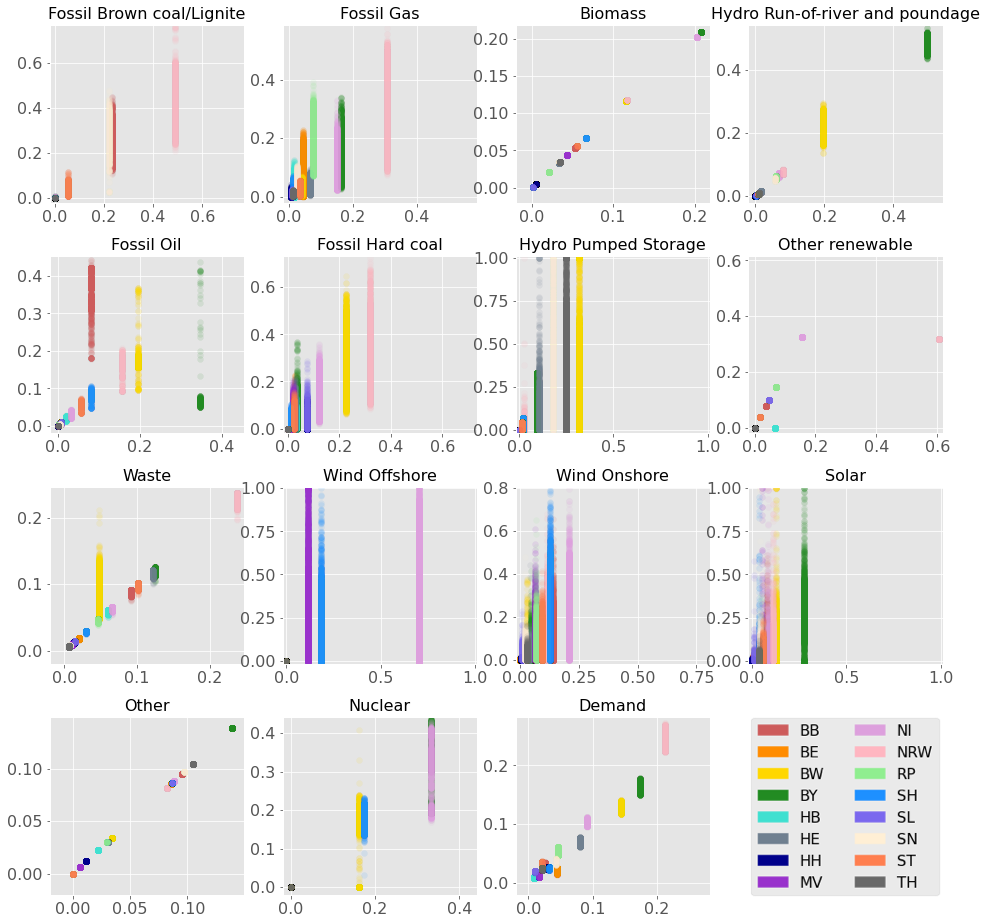}
\caption{Dynamic vs. static regionalization factors for demand and all production types listed in Tab.~\ref{tab:technologies}. For each subfigure, one dot in the scatter plot represents a value for each hour. The x-axis gives the static factor, the y-axis the dynamic factor, different colours indicate different federal states. For some production types and federal states, static and dynamic regionalization factors coincide due to a lack of time-dependent regionalized data.}
\label{fig:vs_everything}
\end{figure*}

A comparison of the static and dynamic regionalization factors for generation and load time series is shown in Fig.~\ref{fig:vs_everything}. The per unit generation dataset does not contain time series for power generation from Biomass, Other, Wind Offshore, Wind Onshore and Solar (see Fig.~\ref{fig:per_generation}). Whereas for Wind Onshore, Wind Offshore and Solar, dynamic regionalization factors can be calculated based on weather data and the distribution of generation capacities, for the remaining not covered generation types no dynamic regionalization factors could be derived. Accordingly, for these technologies, static and dynamic factors coincide in Fig.~\ref{fig:vs_everything}. For the main fossil generation technologies Fossil Brown Coal/Lignite, Hard Coal, Fossil Gas and Nuclear the comparison in Fig.~\ref{fig:vs_everything} indicates the significance of using dynamic regionalization factors. Plant-specific factors like operational costs, heat provisioning in the case of combined heat and power plants, bidding strategies, or outages result in a time-dependent distribution of per type generation which considerably deviates from the distribution of the corresponding generation capacities. This variability also persists for generation from Fossil Gas, where only approximately 50 \% of generation can be regionalized based on per unit generation time series.

The dynamic regionalization factors for Onshore Wind, Offshore Wind and Solar in Fig.~\ref{fig:vs_everything} display an even larger variability as compared to the regionalization factors for fossil generation technologies. Weather patterns can vary considerably over different federal states, resulting in a distribution of per type generation often very different from the distribution of generation capacities. 

For electricity demand, the variability is considerably lower compared to the different production types. This reduced variability can be explained with the overall homogeneous load profiles of households. Nevertheless, in particular, the federal states with larger demand show some significant hourly deviation from the static regionalization factors, mainly caused by industrial loads with different load profiles than households~\cite{kuehnbach2021}.

A major challenge is the validation of the regionalized time series. Generally, regional data is only available with a low temporal resolution (monthly or yearly) through statistical reporting~\cite{LAK}. Different reporting schemes for aggregated statistical data as compared to temporally resolved time series further hinder validation (see for instance the discussion in~\cite{agorameter} or~\cite{Unnewehr2022b}). ENTSO-E publishes generation and load time series for the four different control areas in Germany (operated by the transmission system operators Amprion GmbH, TransnetBW GmbH, TenneT TSO GmbH and 50Hertz Transmission GmbH, respectively), which do not coincide with federal states. To perform a validation using this published data on contral area level, we first derived dynamic regionalization factors for NUTS3 regions using an approach analogous to the one described in Sec.~\ref{sec:dynamic}, and then combined these factors based on the mapping of NUTS3 regions to control areas as provided in~\cite{frysztacki2023}. Applying these dynamic regionalization factors to the published per type generation series for Germany yields regionalized time series for the four control areas, which have been compared with the corresponding time series for these areas as published by ENTSO-E.

Table~\ref{tab:validation} displays the average generation per hour for the main technology types for each control area, both for the published time series and the time series derived from dynamic regionalization factors. Additionally, the root mean square errors between published and derived time series are shown. A very close agreement is observed for Fossil Brown Coal/Lignite and Nuclear. The entire Nuclear power generation is provided by large power plants for which per unit generation time series are available, so all discrepancies result from issues in the original data. Also for Fossil Brown Coal/Lignite, a large share of the generation is covered by published per unit generation time series. Additionally, the generation is almost completely limited to two control areas, further reducing the discrepancy between allocated and actual time series. In contrast, power generation from Fossil Hard Coal and in particular from Fossil Gas also includes a considerable share from smaller power plants distributed over all four control areas. This results in some disagreement between actual and allocated time series  for these technology types, which is even larger for power generation from biomass, for which no per unit generation time series are available. The allocated time series from Solar, Wind Onshore and Wind Offshore are entirely modelled from weather data and information about the distribution of generation capacities. For this approach, the results indicate a good agreement for average generation, with some larger deviations for individual hourly values.

\begin{table*}[htbp]
\centering
\caption{Comparison of regionalized (allocated) and published (actual) time series for German control areas for 2021. First two columns show average generation per technology type or demand per hour in GW (actual and allocated), third column displays the root mean square error between the actual and allocated time series for the given type or demand in GW.}
\label{tab:validation}
\begin{tabular}{l|rrr|rrr|rrr|rrr} 
\hline
ENTSO-E & \multicolumn{3}{c|}{Tennet} & \multicolumn{3}{c|}{50Hertz} & \multicolumn{3}{c|}{TransnetBW} & \multicolumn{3}{c}{Amprion}\\
 technology type & Actual & Allocated & RMSE & Actual & Allocated & RMSE & Actual & Allocated & RMSE & Actual & Allocated & RMSE\\
  \hline\\
Lignite & 0.009 & 0.008 & 0.014 & 5.999 & 6.001 & 0.316 & 0 & 0 & 0 & 5.201 & 5.292 & 0.290\\
Hard Coal & 1.443 & 1.659 & 0.298 & 0.592 & 0.987 & 0.442 & 1.578 & 1.395 & 0.243 & 2.304 & 1.916 & 0.555\\
Gas & 1.541 & 1.927 & 0.563 & 0.369 & 0.926 & 0.622 & 0.232 & 0.362 & 0.174 & 3.840 & 2.766 & 1.189\\
Nuclear & 3.789 & 3.807 & 0.65 & 0 & 0 & 0 & 1.192 & 1.186 & 0.029 & 2.485 & 2.469 & 0.016 \\
Biomass & 2.065 & 1.448 & 0.627 & 0.596 & 1.012 & 0.456 & 0.592 & 1.045 & 0.464 & 1.025 & 0.773 & 0.256\\
Solar & 1.955 & 2.188 & 0.411 & 1.499 & 1.569 & 0.373 & 0.716 & 0.668 & 0.207 & 1.150 & 0.944 & 0.402\\
Wind Onshore & 4.255 & 4.825 & 0.884 & 3.546 & 3.780 & 0.590 & 0.313 & 0.176 & 0.199 & 2.094 & 1.428 & 0.910\\
Wind Offshore & 2.318 & 2.433 & 0.311 & 0.423 & 0.309 & 0.311 & 0 & 0 & 0 & 0 & 0 & 0 \\
Demand & 17.130 & 18.159 & 1.133 & 12.391 & 10.425 & 2.077 &  6.983 & 7.544 & 0.670 & 21.088 & 21.464 & 0.886\\
\hline
\end{tabular}
\end{table*}

\section{Conclusion}\label{sec:conclusion}
Generation and load time series are often publicly available only on a national level. Spatially more detailed information can be estimated based on regionalization factors, which distribute aggregated information to regional level. In this contribution, we have discussed this regionalization process for generation and load time series from national to federal state level for Germany. The necessary time-dependent information can be collected from per unit generation data, load modelling based on load profiles, and renewable generation simulation models leveraging spatio-temporal weather data. We have shown that regionalization factors need to take into account the temporal variability in the distribution of electricity generation and demand. All static and dynamic regionalization factors are available at~\cite{zenodo} to allow immediate application and validation for the scientific community.


\section*{Acknowledgment}
\addcontentsline{toc}{section}{Acknowledgments}
We thank Nick Harder for support with data processing. Madeleine Sundblad acknowledges funding from Elektrizit\"atswerke Sch\"onau (EWS) through the Sonnencent program, ID 00009280. Tim F\"urmann acknowledges funding from DFG (SPP 1984), project ID 450860949.

\bibliographystyle{plain}
\bibliography{references.bib}

\begin{thebibliography}{10}

\bibitem{MaStR}
{Aktuelle Einheiten{\"{u}}bersicht | Marktstammdatenregister}.

\bibitem{BNetzA}
{Bundesnetzagentur Kraftwerksliste}.
\newblock Accessed: 2022-09-13.

\bibitem{electricitymap}
{Electricity Maps}.
\newblock Accessed: 2022-11-30.

\bibitem{ENTSOE_generation}
{ENTSO-E Transparency Platform, Actual Generation per Production Type}.
\newblock Accessed: 2022-11-29.

\bibitem{ENTSOE_load}
{ENTSO-E Transparency Platform, Actual Generation per Production Type}.
\newblock Accessed: 2022-11-29.

\bibitem{SMARD}
Bundesnetzagentur (Federal~Network Agency).
\newblock {SMARD}.
\newblock Accessed: 2022-11-30.

\bibitem{buettner2022}
Clara B{\"u}ttner, Jonathan Amme, Julian Endres, Aadit Malla, Birgit Schachler,
  and Ilka Cu{\ss}mann.
\newblock Open modeling of electricity and heat demand curves for all
  residential buildings in germany.
\newblock {\em Energy Informatics}, 5(1):1--21, 2022.

\bibitem{transparency}
{ENTSO-E}.
\newblock {Transparency Platform}.
\newblock Accessed: 2022-11-30.

\bibitem{energycharts}
{Fraunhofer ISE}.
\newblock {Energy-Charts}.
\newblock Accessed: 2022-11-30.

\bibitem{frysztacki2023}
{Martha Maria} Frysztacki.
\newblock {Mapping of districts to control zones of German Transmission System
  Operators (TSOs)}.
\newblock \url{https://zenodo.org/record/7530196}.

\bibitem{demandregio}
Fabian Gotzens, Bastin Gillessen, Simon Burges, Wilfried Hennings, Joachim
  M{\"u}ller-Kirchenbauer, Stephan Seim, Paul Verwiebe, Tobias Schmid, Fabian
  Jetter, and Timo Limmer.
\newblock {DemandRegio: Harmonisierung und Entwicklung von Verfahren zur
  regionalen und zeitlichen Aufl{\"o}sung von Energienachfragen (Final
  Report)}, 2020.

\bibitem{agorameter}
Fabian Hein and Hauke Hermann.
\newblock {Agorameter -- Dokumentation Version 10}, 2020.

\bibitem{hersbach2020era5}
Hans Hersbach, Bill Bell, Paul Berrisford, Shoji Hirahara, Andr{\'a}s
  Hor{\'a}nyi, Joaqu{\'\i}n Mu{\~n}oz-Sabater, Julien Nicolas, Carole Peubey,
  Raluca Radu, Dinand Schepers, et~al.
\newblock The era5 global reanalysis.
\newblock {\em Quarterly Journal of the Royal Meteorological Society},
  146(730):1999--2049, 2020.

\bibitem{atlite}
Fabian Hofmann, Johannes Hampp, Fabian Neumann, Tom Brown, and Jonas
  H{\"o}rsch.
\newblock {Atlite: a lightweight Python package for calculating renewable power
  potentials and time series}.
\newblock {\em Journal of Open Source Software}, 6(62):3294, 2021.

\bibitem{PyPSA-Eur}
Jonas H{\"o}rsch, Fabian Hofmann, David Schlachtberger, and Tom Brown.
\newblock {PyPSA-Eur: An open optimisation model of the European transmission
  system}.
\newblock {\em Energy Strategy Reviews}, 22:207--215, 2018.

\bibitem{jung2022}
Christopher Jung and Dirk Schindler.
\newblock On the influence of wind speed model resolution on the global
  technical wind energy potential.
\newblock {\em Renewable and Sustainable Energy Reviews}, 156:112001, 2022.

\bibitem{kuehnbach2021}
Matthias K{\"u}hnbach, Anke Bekk, and Anke Weidlich.
\newblock {Prepared for regional self-supply? On the regional fit of
  electricity demand and supply in Germany}.
\newblock {\em Energy Strategy Reviews}, 34:100609, 2021.

\bibitem{LAK}
{L\"anderarbeitskreis Energiebilanzen}.
\newblock {Energiebilanzen}.
\newblock Accessed: 2022-11-30.

\bibitem{energieatlas}
{Landesamt für Natur, Umwelt und Verbraucherschutz Nordrhein-Westfalen}.
\newblock {Methodik zum Strommarktmonitoring NRW}, 2022.

\bibitem{risch2022}
Stanley Risch, Rachel Maier, Junsong Du, Noah Pflugradt, Peter Stenzel, Leander
  Kotzur, and Detlef Stolten.
\newblock Potentials of renewable energy sources in germany and the influence
  of land use datasets.
\newblock {\em Energies}, 15(15):5536, 2022.

\bibitem{zenodo}
Madeleine Sundblad, Tim F\"urmann, Anke Weidlich, and Mirko Sch\"afer.
\newblock {Load and generation time series for German federal states: Static
  vs. dynamic regionalization factors (data)}.
\newblock \url{http://zenodo.org/record/7510855}.

\bibitem{tranberg}
Bo~Tranberg, Olivier Corradi, Bruno Lajoie, Thomas Gibon, Iain Staffell, and
  Gorm~Bruun Andresen.
\newblock {Real-time carbon accounting method for the European electricity
  markets}.
\newblock {\em Energy Strategy Reviews}, 2019.

\bibitem{Unnewehr2022}
Jan~Frederick Unnewehr, Mirko Sch\"{a}fer, and Anke Weidlich.
\newblock {The value of network resolution – A validation study of the
  European energy system model PyPSA-Eur}.
\newblock In {\em 2022 Open Source Modelling and Simulation of Energy Systems
  (OSMSES)}, pages 1--7, 2022.

\bibitem{Unnewehr2022b}
Jan~Frederick Unnewehr, Anke Weidlich, Leonhard Gf{\"u}llner, and Mirko
  Sch{\"a}fer.
\newblock Open-data based carbon emission intensity signals for electricity
  generation in european countries--top down vs. bottom up approach.
\newblock {\em Cleaner Energy Systems}, 3:100018, 2022.

\bibitem{NaturalEarth}
Nathaniel Vaughn~Kelso and Tom Patterson.
\newblock {Natural Earth Data}.
\newblock Accessed: 2022-11-20.

\bibitem{victoria2019}
Marta Victoria and Gorm~Bruun Andresen.
\newblock Using validated reanalysis data to investigate the impact of the pv
  system configurations at high penetration levels in european countries.
\newblock {\em Progress in Photovoltaics: Research and Applications},
  27(7):576--592, 2019.

\bibitem{WindGuard}
Deutsche WindGuard.
\newblock {Status des Offshore-Windenergieausbaus in Deutschland Jahr 2020}.
\newblock Accessed: 2022-09-25.

\end{thebibliography}

\end{document}